\documentstyle[twoside,fleqn,npb,epsfig]{article}

\newcommand{\AmS}{{\protect\the\textfont2
  A\kern-.1667em\lower.5ex\hbox{M}\kern-.125emS}}

\newcommand{\zf}{{\tt ZFITTER}} 
\newcommand{\nn}{\nonumber} 
 
\newcommand{\bq}{\begin{equation}} 
\newcommand{\eq}{\end{equation}} 
\newcommand{\ba}{\begin{eqnarray}} 
\newcommand{\ea}{\end{eqnarray}}

\newcommand{\prbr}{\scriptstyle (\prime)}
\newcommand{\mz}{M_{_Z}}
\newcommand{\afba}[1]{A^{#1}_{_{\rm FB}}}
\newcommand{\re}{\Re e}

\title{Predictions for Fermion Pair Production at $e^+ e^-$ Colliders}
\author{\underline{M. Jack}\address{Deutsches Elektronen-Synchrotron DESY,
Platanenallee 6, 
D-15738 Zeuthen, Germany }
\thanks{Talk given at ``Loops and Legs 2000'', Bastei, Germany, April 9-14,
to appear in the Proceedings.},
        A. Hoefer$^a$, A. Leike\address{Ludwig-Maximilians-Universit\"at, 
        Sektion Physik, 
Theresienstr. 37,
D-80333 M\"unchen, Germany}, 
        and T. Riemann$^a$
\vspace{-4cm}
\begin{flushleft}
{\tt
DESY 00-073
\\
hep-ph/0007046
}
\end{flushleft}
\vspace{3cm}
}

\begin{document}

\begin{abstract}
We discuss the status and some ongoing upgrades of the 
{\tt ZFITTER} program for applications at $e^+e^-$ colliders
LEP~1/SLC, LEP~2, GigaZ, and TESLA. The inclusion of top quark pair
production is under work.
\end{abstract}
\maketitle

\setcounter{footnote}{3}
\renewcommand{\thefootnote}{\fnsymbol{footnote}}

\section{Introduction}
%
One of the great achievements in theoretical and 
experimental physics over the last three decades  
is the unique calculation of 
quantum corrections in the Standard Model ({\tt SM})
of fundamental particle interactions,  
following the pioneering work of G. 't~Hooft and M.~Veltman
\cite{Veltman:1968ki,'tHooft:1971rn,'tHooft:1972fi}, and their
observational evidence at the 10 $\sigma$ confidence level
\cite{Sirlin:1999zc}.  
Especially the electroweak sector of the {\tt SM} is being
investigated at the high energy $e^+e^-$ colliding facilities 
LEP~1 and SLC, and now at LEP 2, with unprecedented precision.
This, though, could well be surpassed in the future e.g.~by the 
GigaZ mode of the TESLA Linear Collider (LC) project
\cite{Moenig:1999,Erler:2000}.  
One main focus, besides the search for direct signals of New Physics, 
was and still is on studying the 
properties of the neutral and charged weak gauge bosons. 

Here we concentrate on the measurement of different 
cross sections and asymmetries in fermion pair
production. Fermion pair production is
an essential tool to determine parameters
like the mass and total and partial
decay widths of the weak neutral gauge boson 
$Z$ and the neutral current fermionic 
couplings.
The theoretical break-through for this was the first 
complete evaluation of electroweak quantum corrections
to $e^+e^-\to \mu^+ \mu^-$ in the Weinberg model in \cite{Passarino:1979jh},
with a collection of the scalar one-loop integrals first
given in \cite{'tHooft:1979xw}. 

For today's predictions at the few per mil level,
this calculation can only be seen as the first essential
step. It did not include such important ingredients as a 
refined treatment of the $Z$ boson resonance, of QCD and 
higher order corrections, or realistic QED hard bremsstrahlung effects. 

During the last twenty years many dedicated efforts were invested to
perform the necessary improvements, to implement them into computer 
codes which may be applied in the analysis of experimental data and
in regularly repeated comparisons of the codes.
A representative collection of the underlying expressions, their
implementations and numerical comparisons may be found in 
\cite{Bardin:1989qr,Bohm:1989pb,Bardin:1995aa,Boudjema:1996qg};
see also
\cite{Beenakker:1997fi,Bardin:1998nm,Bardin:1999gt,Christova:1999gh} 
for individual comparisons.
The most recent workshop on LEP~2 physics is yet under way
\cite{Was:2000??}.
Prominent examples of such codes are: 
\begin{itemize}
\item[${-}$]
{\tt ALIBABA},
\vspace{-2mm}
\item[${-}$]
{\tt BHM},
\vspace{-2mm}
\item[${-}$]
{\tt KORALZ}, {\tt KK},
\vspace{-2mm}
\item[${-}$]
{\tt TOPAZ0},
\vspace{-2mm}
\item[${-}$]
{\tt ZFITTER}. 
\end{itemize}

On the latter, the semi-analytical Fortran program {\tt ZFITTER} 
\cite{Bardin:1999yd-orig},
we will now focus. 
Earlier program descriptions are \cite{Bardin:1989tq,Bardin:1992jc2}.
The core of \zf\ relies on a complete electroweak one-loop calculation 
\cite{Bardin:1980,Bardin:1989di,Bardin:1991fu,Bardin:1991de,Christova:1999cc}.
It is also based on many additional formulae; from
\cite{Bardin:1999yd-orig} one may extract a list of papers that \zf\ uses in addition,  
\cite{%
Akhundov:1986fc,%
Arbuzov:1995id,%
Arbuzov:1999uq,%
Arbuzov:1999uq1,%
Arbuzov:1992pr,%
Avdeev:1994db,%
Barbieri:1993ra,%
Bardin:1986fi,%
Bardin:1987hva,%
Bardin:1988xt,%
Bardin:1989aa,%
Bardin:1989cw,%
Bardin:1991xe,%
Bardin:1999ak,%
Beenakker:1989km,%
Berends:1988ab,%
Berman:1958,%
Bilenkii:1989zg,%
Bohm:2000jw,%
Bonneau:1971mk,%
Celmaster:1980xr,%
Chetyrkin:1979bj,%
Chetyrkin:1994js3,%
Degrassi:1996mg,%
Degrassi:1996ZZ,%
Degrassi:1997ps,%
Degrassi:1999jd,%
Dine:1979qh,%
Djouadi:1987gn,%
Djouadi:1988di,%
Eidelman:1995ny,%
Field:1996dk,%
Fleischer:1992fq,%
Gorishnii:1991hw,%
Greco:1975,%
Harlander:1998zb,%
Jadach:1992aa,%
Jegerlehner:1995ZZ,%
Kallen:1955ks,%
Kallen:1968,%
Kataev:1992dg,%
Kinoshita:1959ru,%
Kirsch:1995cf1,%
Kirsch:1995cf,%
Kniehl:1988id,%
Kniehl:1990yc1,%
Kniehl:1990yc,%
Kuraev:1985hb,%
Leike:1991a,%
Leike:1991pq,%
Leike:1992uf,%
Mann:1984dv,%
Matsuura:1987,%
Montagna:1989,%
Montagna:1997jv,%
Passarino:1982,%
Riemann:1997aa,%
Riemann:1992gv,%
Riemann:1997kt,%
Riemann:1997tj1,%
Sedykh:19xy,%
Silin:19xy,%
Skrzypek:1992vk,%
Steinhauser:1998rq,%
Stuart:1991xk,%
vanderBij:1984bw,%
vanderBij:1984aj,%
vanderBij:1987hy,%
vanRitbergen:1998%
}.

\section{Status of two-fermion codes at LEP and higher energies} 
%
{\tt ZFITTER} \cite{Bardin:1999yd-orig} 
calculates radiative corrections to the muon decay constant
\cite{Bardin:1980,Akhundov:1986fc,Bardin:1987hva,Bardin:1995aa}, 
to the $Z$ decay width 
\cite{Akhundov:1986fc,Bardin:1987hva,Bardin:1995aa},
and to the $W$ decay width \cite{Bardin:1986fi}\footnote{
A Fortran bug was corrected in {\tt ZFITTER} v.6.30 
\cite{Bardin:1999yd-orig} resulting in a 0.3\% 
shift of the $W$ partial widths.}.   
Cross sections and asymmetries are treated in a semi-analytical 
approach. Improved Born observables $\sigma^0_{T,FB}(s^{\prbr},s^{\prbr})$ 
\cite{Bardin:1987hva,Bardin:1988xt,Bardin:1989di,Bardin:1995aa}  
containing the complete virtual weak and QCD corrections
are convoluted with analytical flux functions $\rho(s'/s)$ 
(radiators) for the QED corrections
\cite{Bardin:1989cw,Bardin:1991fu,Bardin:1991de,Christova:1999cc}. 
This is done in a one-dimensional numerical integration over the
final-state invariant mass squared  
$s' = M_{f\bar{f}}^2$ \cite{Bardin:1999yd-orig}.
The especial Bhabha scattering case is treated in an effective Born
approximation only following \cite{Bardin:1995aa,Bardin:1991xe}. 
As an example, the initial-state corrections to total 
cross sections for $s$-channel processes are reproduced in short ($v \equiv 1 - {s'}/s$):
\ba
\label{eq:1}
\sigma^{ini}_T(s) 
\hspace*{-2.5mm}
&=&  
\hspace*{-2.5mm}
\int d\Biggl(\frac{s'}{s}\Biggr)~ 
\sigma^0_T(s')~\rho^{ini}_T\Biggl(\frac{s'}{s}\Biggr),
\\  
\label{eq:2}
\rho^{ini}_T\left(\frac{s'}{s}\right) 
\hspace*{-2.5mm}
&=&  
\hspace*{-2.5mm}
\left(1+{\bar S}^{ini}\right)\beta_e v^{\beta_e-1} 
+ {\bar H}_{T}^{ini}\left(\frac{s'}{s}\right),
\\
\beta_e 
\hspace*{-2.5mm}
&=&  
\hspace*{-2.5mm}
\frac{2\alpha}{\pi} Q_e^2 \left( \ln \frac{s}{m_e^2}-1\right).
\ea
The radiators $\rho^{ini}_T$, $\rho^{ini+fin}_T$, and $\rho^{int}_T$
with ${\bar S}$ and ${\bar H}_T$ 
are determined in \cite{Bardin:1989cw}, and    
including all relevant higher order
terms e.g.~in \cite{Bardin:1999yd-orig}.   
Eq.~(\ref{eq:1}) can be straightforwardly 
generalized to different asymmetries $A_{FB}$, $A_{\rm pol}$,
$A_{LR}$ etc. or to scattering angle distributions like  
$d{\sigma}/{d{\cos\theta}}$, then with different 
effective Born terms and radiators. Kinematical cuts to the 
final-state phase space may also be applied 
\cite{Passarino:1982,Bardin:1991fu,Bardin:1991de,Christova:1999cc,%
Montagna:1993mf}.
Also a more model-independent description of cross section 
observables, e.g. in form of an S-matrix approach, 
can be thought of 
\cite{Leike:1991pq,Stuart:1991xk,Kirsch:1995cf1,Kirsch:1995cf,%
Riemann:1992gv,Riemann:1997kt,Riemann:1997tj1,Bohm:2000jw}.

Concerning the status of present codes for LEP~1 and SLC 
applications, one can summarize that the level of precision 
for cross section predictions in $s$-channel fermion pair 
production is now better than $10^{-4}$ on the $Z$ boson 
resonance and better than $0.3\times 10^{-3}$ for center-of-mass 
energies $\sqrt{s}=M_Z\pm 3$ GeV (single contributions). 
For the recently 
updated branch with cuts on final-state maximum acollinearity 
and minimum energies in {\tt ZFITTER}
\cite{Christova:1999gh,Christova:1999cc}, this is demonstrated in 
Table~\ref{tab:acol}, comparing with program {\tt TOPAZ0}
\cite{topaz0} when the QED initial-final state interference is switched on.
%
\begin{table*}[hbt]
\setlength{\tabcolsep}{1.5pc}
\newlength{\digitwidth} \settowidth{\digitwidth}{\rm 0}
\catcode`?=\active \def?{\kern\digitwidth}
\caption[]{
{A comparison of predictions from {\tt ZFITTER} v.6.30 
\cite{Bardin:1999yd-orig} (Jun 2000)
and {\tt TOPAZ0} v.4.4 \cite{topaz0}
for muonic cross sections and forward-backward
asymmetries around the $Z$ peak.
First row is without initial-final state interference, second row with,
third row the relative/absolute effect of that interference in per mil
($M_Z=91.1871\pm 0.0021$ GeV, $M_H=125$ GeV, $m_t=173.8$ GeV,
$\alpha_S=0.119$, $\Delta\alpha_{\rm had}^{(5)}(M_Z)=0.0280398089$).
}
}
\label{tab:acol}
\begin{tabular*}
{\textwidth}
{c||c|c|c|c|c}
\hline
\multicolumn{6}{c}{
{$\sigma_{\mu}\,$[nb] with $\theta_{\rm acol}<10^\circ$}}
\\ 
\hline
$\theta_{\rm acc} = 0^\circ$& $\mz - 3$ & $\mz - 1.8$ & $\mz$ & $\mz + 1.8$ &
$\mz + 3$  
\\ 
\hline\hline
  & 0.21928  & 0.46282  & 1.44814  & 0.67722  & 0.39362 
\\
{{\tt TOPAZ0}}  & 0.21772  & 0.46077  & 1.44805  & 0.67891  & 0.39486 
\\
 & {\bf --7.17}    
 & {\bf --4.45}     
 & {\bf --0.06}     
 & {\bf +2.49}     
 & {\bf +3.14}    
\\ 
\hline
& 0.21923  & 0.46278  & 1.44794  & 0.67716  & 0.39356\\
{{\tt ZFITTER}}  & 0.21768 & 0.46075  & 1.44790  & 0.67893  & 0.39485 \\ 
  & {\bf --7.16}     
  & {\bf --4.41}     
  & {\bf --0.03}    
  & {\bf +2.61}
  & {\bf +3.27}   
\\ 
\hline 
\hline
\multicolumn{6}{c}{$\afba{\mu}$ with $\theta_{\rm acol}<10^\circ$} \\
\hline
$\theta_{\rm acc} = 0^\circ $& $\mz - 3$ & $\mz - 1.8$ &
$\mz$ & $\mz + 1.8$ & $\mz + 3$  \\ 
\hline\hline
  & --0.28473 & --0.16935  & 0.00014  & 0.11494  & 0.16089 \\
{{\tt TOPAZ0}}  & --0.28181 & --0.16686  & 0.00068  & 0.11367  & 0.15919 \\ 
  & {\bf +2.92}    
  & {\bf +2.49}     
  & {\bf +0.54}     
  & {\bf --1.27}
  & {\bf --1.70}    
\\ 
\hline
  & --0.28519 & --0.16958  & 0.00005  & 0.11479  & 0.16068 \\
{{\tt ZFITTER}}  & --0.28244 & --0.16731  & 0.00065  & 0.11375  & 0.15909\\
 & {\bf +2.75}    
 & {\bf +2.27}     
 & {\bf +0.60}    
 & {\bf --1.04}
 & {\bf --1.59}
\\
\hline 
\end{tabular*}
\end{table*}
%
For earlier comparisons with acollinearity cuts also consult 
\cite{Christova:1999gh,Christova:1998tc,Jack:1999xc,Jack:1999af}. 

At LEP 2 center-of-mass energies up to roughly 200 GeV 
an agreement of the codes at the order of few per mil is being
achieved \cite{Was:2000??}.
This was studied earlier for the two-fermion programs {\tt ZFITTER} 
\cite{Bardin:1992jc2,Bardin:1999yd-orig},
{\tt BHM} \cite{Burgers:BHM,Bardin:1995aa},
{\tt TOPAZ0} \cite{topaz0}, and {\tt KORALZ/KK} \cite{koralzkk,Jadach:1999kkkz}
for different cuts on minimum $s'$ 
\cite{Bardin:1995aa,Montagna:1997jt,Bardin:1998nm,Bardin:1999gt,Jadach:1999in}
or on maximum acollinearity and minimum energies of the final-state fermions
\cite{Christova:1999gh,Christova:1998tc,Jack:1999xc,Jack:1999af}. 
Similar studies of Bhabha
scattering include the code {\tt ALIBABA} \cite{Beenakker:1991mb} 
in \cite{Beenakker:1997fi,Jadach:2000ll}.
The conclusion was that precisions are of the order of few 
per mil to $1\%$ for LEP 1 or LEP 2 energies respectively, excluding 
a radiative return to the $Z$ by sufficiently strong cuts. 
Compared 
with the final experimental precisions at LEP 1/SLC and LEP 2 these 
theoretical accuracies are satisfactory \cite{Quast:2000ll}. 
It is worth mentioning that for the envisaged accuracy several recent 
improvements had to be undertaken, e.g. a better treatment of initial-
and final-state pair production corrections, 
the exponentiation of initial-final state corrections, convolution of
the $ZZ$ and $WW$ box corrections, etc. 
As a result, {\tt ZFITTER} v.6.30 \cite{Bardin:1999yd-orig} 
became more complicated and also
slower, and the same might be true also for the 
other codes.

\begin{figure}[htb]
\vspace{9pt}
\framebox[75mm]{\rule[75mm]{5mm}{-5mm}
\mbox{
\epsfig{file=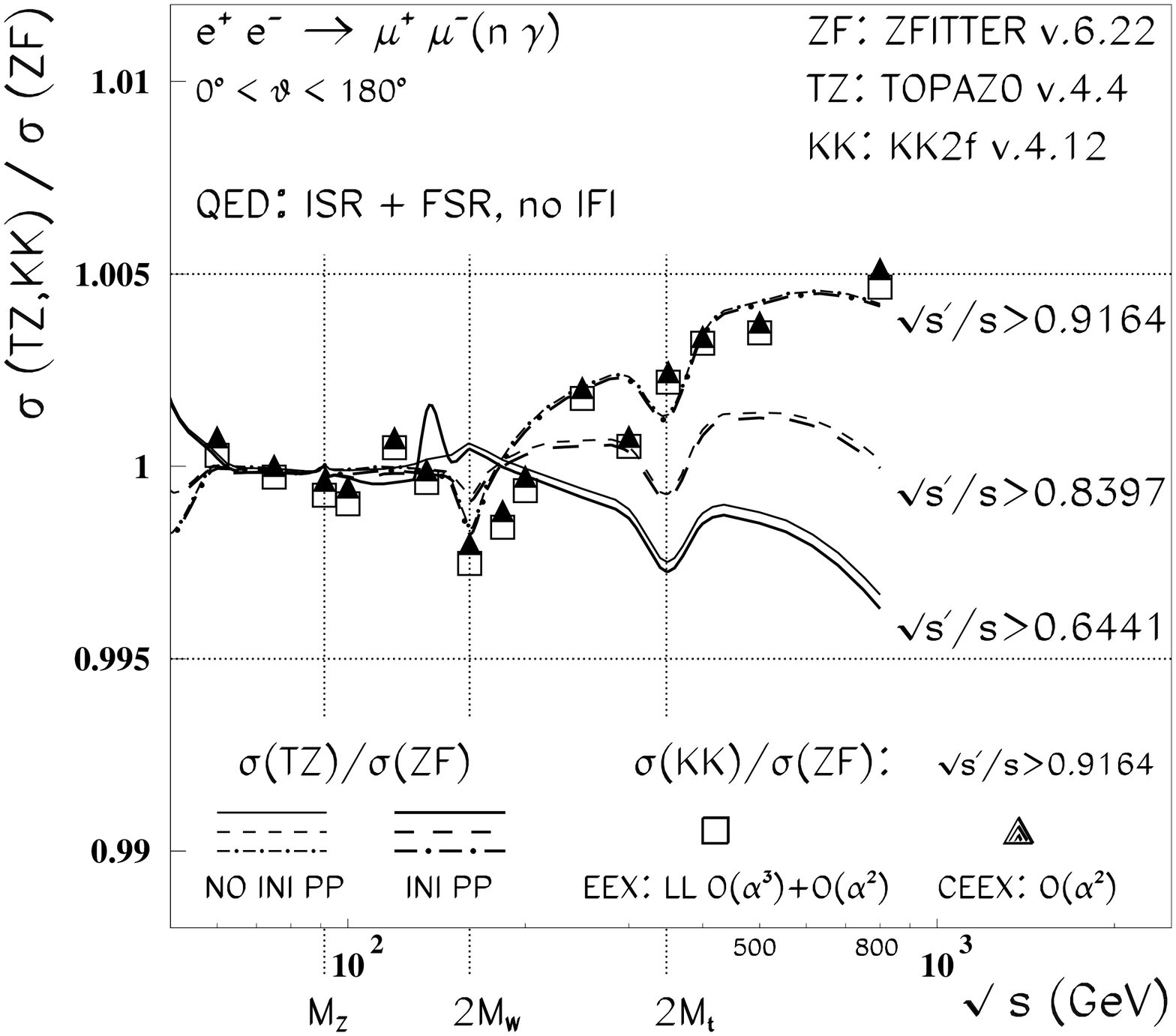,width=8cm}}
}
\caption{
Cross section ratios for muon-pair production with different $s'$ cuts
for codes {\tt ZFITTER} v.6.22 (Oct 1999) \cite{Bardin:1999yd-orig}, 
{\tt TOPAZ0} v.4.4 \cite{topaz0}, 
{\tt KK} v.4.12 \cite{Jadach:1999kkkz} from 60 to 800 GeV c.m.~energy;
without initial-final state interference
({\tt INI PP}: initial-state pair production; {\tt LL}: leading logarithmic terms)
\cite{Christova:2000zu}.
}
\label{fig:codes}
\end{figure}

Thinking of applications of the {\tt ZFITTER} code 
and of other programs at a future $e^+e^-$ Linear Collider (LC)
with much higher luminosities and energies, the above 
observations at LEP/SLC energies might only prove to be 
preliminary and a further upgrade could become necessary. 
Two different scenarios may be envisaged
for the use of the {\tt ZFITTER} code at a LC:
 
First, precision physics 
could again be performed on the $Z$ boson resonance, but 
then with 1000 times the luminosity of LEP~1
\cite{Moenig:1999,Erler:2000} 
in order to search e.g.~for virtual effects from a {\tt SM} 
or {\tt MSSM} (minimal supersymmetric) Higgs boson or from 
supersymmetric particles \cite{Heinemeyer:1999aa,Erler:2000}. 
With such a GigaZ option, experimental accuracies
could increase by a factor of 100 or more \cite{Moenig:1999,Erler:2000}. 
Updates of the codes 
could include higher order QED radiative  effects,
effects by beamstrahlung, or an update of the still critical
Bhabha scattering case when demanding high precisions.
What is more, complete electroweak two loop calculations 
\cite{Freitas:2000ll} might become necessary; see also the 
many talks on this topic at this conference.   
At energies up to roughly 800 GeV like for the {\it Tesla project} 
\cite{Brinkmann:1997nb}, issues like experimental and 
theoretical precisions are still quite vague, but it is clear that 
the demands on codes for two-fermion production will be quite
challenging due to  
higher beam energies, higher luminosities, and improved analysis 
techniques. On the one hand, 
electroweak and QED corrections become equally 
important which may demand a critical look at the numerical 
validity of the usually applied improved Born approximation 
at higher energies. 
On the other hand, higher order 
electroweak corrections will also grow in importance with 
increasing energies \cite{sudakov}. 
First studies of codes 
{\tt ZFITTER}, {\tt TOPAZ0}, and {\tt KK} show some evidence that an 
agreement of 5 per mil to 1 per cent can be reached 
for the complete energy range \cite{Christova:2000zu,LEP2MCWS2000}
(see Fig.~\ref{fig:codes}).
 
\section{Top pair production at LC energies
\label{sub_out_ttbar}
}
%
At a future LC, top pair production will be 
studied in detail.
One of the main topics of investigation 
will be the analysis of the elementary properties 
of the top quark,
including 
the measurement of its mass $m_t$, total and 
partial decay widths, and couplings to the weak gauge bosons.
The recent progress
in calculations at the $t\bar{t}$ {\em threshold} itself
is summarized e.g. in \cite{Hoang:2000yr} and is continuously being
updated in \cite{topreport}.  
A determination of the weak neutral vector and axial-vector
couplings of the top quark to the $Z$ boson 
will be performed in the perturbative region sufficiently 
above $\sqrt{s}\approx 350\,\mbox{GeV}$
by measuring cross sections, forward-backward and polarization
asymmetries.
 Here {\tt ZFITTER} may naturally step in to do the job. 
The one key point now to be considered is the inclusion 
of the mass of the final-state heavy fermions. Up to now 
final-state masses could be neglected 
for applications at the LEP or SLC center-of-mass energies.

For the massive case, the complete one-loop electroweak corrections 
in the {\tt SM} with the running of the QED coupling, fermionic 
self energies, vertex corrections, and electroweak box contributions 
have been determined in \cite{Beenakker:1991ca}. 
A recent update of the QCD effects was given in 
\cite{Ravindran:1998jw}. 
Work is right now in progress treating 
the electroweak effects in form of effective
couplings together with form factors in an  
improved Born approach [see term $\sigma^0_T(s')$ in 
(\ref{eq:1})], suitable for a quick calculation of interesting 
observables with the {\tt ZFITTER} code \cite{BKNRL:1999up}. 
For results within the {\tt MSSM} please refer to 
\cite{Hollik:1999md,Guasch:1998kt}.

In the remaining part of this contribution, we focus on the
top mass dependent radiative  
corrections from QED bremsstrahlung which correspond to the 
radiators $\rho^a_A(s'/s)$, $a=ini,fin,int$, $A=T,FB$ [see (\ref{eq:1})]. 
One may expect at most some additional several per cent photonic 
corrections from the finite top quark mass. 
Initial-state corrections to $\sigma_T$ \cite{Akhundov:1991qa} 
and final-state effects to the angular distribution
$d\sigma/d\cos\vartheta$ (i.e. also to $A_{FB}$) are known 
\cite{afbmass,Arbuzov:1992pr,Ravindran:1998jw}. 

For the massive cross sections
$\sigma_T, A_{FB}=\sigma_T/\sigma_{F-B}, d\sigma/d{\cos\vartheta}$, 
analytic expressions like (\ref{eq:1}) would 
be nice to have. 
Soft and virtual QED corrections with final-state masses have been 
known for a long time for pure QED \cite{Berends:1983dy}
and also in the {\tt SM} \cite{Beenakker:1991ca,Hollik:1999md}. 
For the hard massive radiators $\rho^a_A(s'/s;m_f^2/s)$ 
the situation is the following:
With an $s'$ cut only, hard photonic corrections 
for total cross sections $\sigma_T(s)$ with $m_f\neq 0$ were
shown in \cite{Akhundov:1991qa} and \cite{Arbuzov:1992pr}
for the initial- and final-state corrections respectively. 
Having in mind a cut on the cosine of the scattering angle 
$\cos\vartheta$ of
one final-state fermion, the differential cross section
$d{\sigma}/d{\cos\theta}$ first has to be determined. 
For this case, only 
formulae with final-state bremsstrahlung
are given in \cite{Arbuzov:1992pr};
without cuts,  the total cross section and 
forward-backward asymmetry had been presented 
earlier in \cite{afbmass}.

It should be mentioned that the QED corrections to cross 
section contributions $\sigma_L$ and $\sigma_R$ for a left-
or right-handed polarized $e^-$ beam are not automatically given 
by the results $\rho^a_A$ for unpolarized cross sections; 
there are additional  
radiators that do depend on the initial helicities \cite{QEDM}. 
Of course, the effective 
Born results $\sigma^0_T$ also have to be replaced 
by the corresponding massive terms $\sigma_{L,R}^0$
with polarization. 
Nevertheless, the predictions for  $A_{LR}$, as being 
measured at SLC at the $Z$ resonance using a polarized $e^-$ beam, 
may be to a very good precision described by the massless radiator functions 
(as is presently done in {\tt ZFITTER}) since there hard bremsstrahlung 
is strongly suppressed, and, in addition, for polarization asymmetries the 
photonic corrections cancel to a large extent 
(see e.g. \cite{Riemann:1992gv}).

\subsection{\label{sec-ana}The distribution $d\sigma_T^{int}/ds'$
} 
%
The hard-photon contribution from QED interference 
to total cross sections for heavy fermions,
\ba
\label{reac}
e^-(p_1) + e^+(p_2) \to f(p_3) + \bar{f}(p_4) + \gamma(k),
\ea 
will be calculated here by two methods; one is 
tensor integration.
The final-state phase space is split 
into two two-particle phase spaces.
The $(f\bar{f})$ 
rest frame is boosted with respect to the center-of-mass system (cms).
The phase space integration is then conveniently 
carried out in two steps:
First, the final-state tensor ${\cal F}^{\mu\nu}$ is integrated over
$p_3$ and $p_4$ in the $(f\bar{f})$ rest frame,
leaving Lorentz covariant expressions in 
$k^\rho$ and $q^\rho \equiv p_3^\rho + p_4^\rho$. 
Contracting then with the initial-state tensor ${\cal I}_{\mu\nu}$,
the remaining integration over the cosine of the photon 
polar angle with respect to the beam axis is 
carried out in the center-of-mass system. 
Evidently, the cms fermion production angle is not 
accessible in this approach and thus neither the angular distribution nor 
$A_{FB}$. 
The integrated distribution gets:
\begin{eqnarray}
  \label{eq:defint}
  \frac{d\sigma_T^{int}}{ds'} = \sum\limits_{V_{1,2}=\gamma, Z, Z'}
\frac{d\sigma_T^{int}}{ds'}(V_1 V_2),
\end{eqnarray}
where
\ba
\label{eq:zprz}
\frac{d\sigma_T^{int}}{ds'}(Z'Z)
\hspace*{-0.2cm}&=&\hspace*{-0.2cm}
-4\;\alpha^3\;Q_e Q_f\; 
\\
&&\hspace*{-0.2cm}\cdot~
\re[\chi_{Z'}(s')\chi_Z^*(s)]
( v_e a'_e + a_e v'_e ) 
\nn\\
&&\hspace*{-0.2cm}\cdot~
\frac{1}{s^2}\frac{ 1 + s'/s }{ 1 - s'/s }
\Bigl[
\beta_f( v_f a'_f + a_f v'_f )
\nn\\
&&\hspace*{-0.2cm}
-~\frac{2 m_f^2}{s} L_f 
\left( \frac{s}{s'} v_f a'_f + a_f v'_f \right)
\Bigr], 
\nn
\ea
and
\ba
\label{eq:lf} 
L_f &=& \log\left(\frac{1+\beta_f}{1-\beta_f}\right),
\\
\label{eq:betaf} 
\beta_f &=& \sqrt{1-\frac{4 m_f^2}{s'}},
\\ 
\label{eq:deltaz}
\chi_Z(s) &=& \frac{G_{\mu} M_Z^2}{\sqrt{2}~2\pi\alpha}
\frac{s}{s - M_Z^2 + i M_Z \Gamma_Z}.
\ea
%
For $Z$-boson exchange we have:
$v_{f} = I_3^{f} - 2 Q_{f}\sin^2\theta_W$
and $a_{f} = I_3^{f}$ with $I_3^e = -1/2$, $Q_e = -1$.
For the contributions from $\gamma$-exchange, for example, 
we have simply $v_f = Q_f$, $a_f = 0$, etc. and 
$\chi_{\gamma} \to 1$.
Massive contributions are proportional to $2 m_f^2/s\cdot L_f$
which drop out for $\beta_f\to 1$ in the massless case.
The formula (\ref{eq:zprz}) is quite 
compact.

It may of course 
also be used in the context of searches for extra
heavy gauge bosons $Z'$ e.g.~through $Z Z'$ mixing effects
(see e.g.~\cite{Leike:1991a,Leike:1992uf,Djouadi:1992sx} and program
{\tt ZEFIT} \cite{Riemann:1997aa}).
The massless limit is given in Eq. (1.3.14) in \cite{Bardin:1999yd-orig}.

Numerical results are shown in Fig.~\ref{fig:dsig1}.
We obtained that figure also with another approach (see the next section), 
where we also include a short discussion.

\subsection{\label{sec-num}The distribution $d\sigma_{FB}^{int}/ds'$
}
Another phase space parameterization makes the fermion 
production angle accessible.
The approach goes back to 
\cite{Passarino:1982} and is
explained in
sections (1.2) and (1.5.1) of \cite{Bardin:1999yd-orig}; 
see also \cite{Christova:1999gh} for more details, where also the first 
analytical integration is described.
After that, the integration is much simplified, and it is here where presently 
the Fortran program {\tt topfit.f} \cite{QEDM}
starts the numerical treatment.
We have some understanding of the resulting integrand, and it seems to us 
that the remaining analytical integrations may also be performed 
while retaining a finite 
final fermion mass $m_f$ \cite{QEDM}.
This would open the way to treat the massive 
initial-final state interference on the 
same footing as the other corrections, 
namely with only one numerical 
integration (over $s'$).

\begin{figure}[htb]
\vspace{9pt}
\framebox[75mm]{\rule[75mm]{5mm}{-5mm}
\mbox{
\epsfig{file=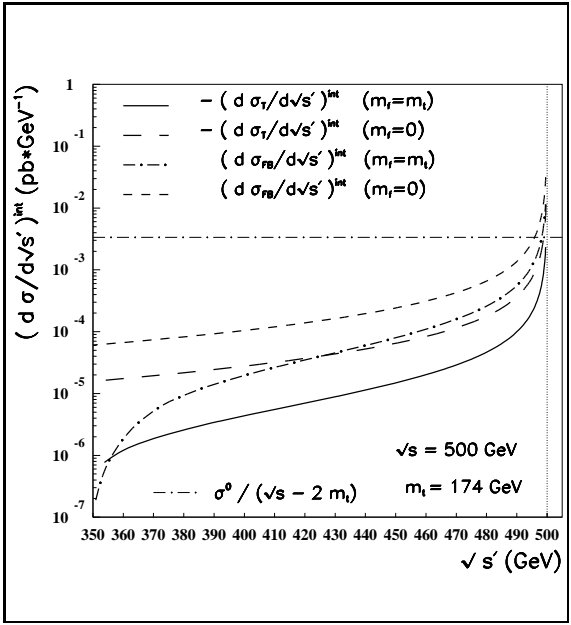,width=8cm}}
}
\caption
{Invariant mass cross section distributions 
${d\sigma_{T}^{int}}/{d\sqrt{s'}}$
and ${d\sigma_{FB}^{int}}/{d\sqrt{s'}}$.
}
\label{fig:dsig1}
\end{figure}

The numerical effects from final 
mass corrections to the initial-final state interference QED bremstrahlung 
are depicted in 
Fig.~\ref{fig:dsig1}. 
There, both the invariant mass distributions 
${d\sigma_{T,FB}^{int}}/{d\sqrt{s'}}$ 
are shown
without and with final-state mass effects for $m_f = m_t = 174$ GeV and 
$\sqrt{s} = 500$ GeV.
The tree-level 
cross section, shown for comparison, 
is $\sigma^0=0.51$ pb. 

For large invariant masses, $\sqrt{s'}\to \sqrt{s}$, the 
infrared peak clearly dominates.
It has to be regularized by soft and 
virtual corrections. 
The main difference
between the massive and massless cases is a suppression of 
cross sections near the threshold $\sqrt{s'}\approx 2 m_t$. 
This suppression is stronger for $d\sigma_{FB}^{int}$
than for $d\sigma_{T}^{int}$.
The total corrections from bremsstrahlung are roughly by a factor 
four larger for $d\sigma_{FB}$ than for $d\sigma_{T}$
in both cases. 
Including final-state masses $m_t$, however,
reduces the radiative corrections for both $d\sigma_{FB}$ and $d\sigma_{T}$
again by roughly a factor of four. 
\section{Conclusions}
%
Summarizing, the semi-analytical Fortran program 
{\tt ZFITTER} for $e^+e^-\to f\bar{f}$ with radiative 
corrections is applied in electroweak precision physics at LEP and SLC.
With its quick cross section evaluations and precision of 
about $10^{-3}$ at the $Z$ peak and of order 
few per mil to $1\%$ up to $\sqrt{s}\approx 800$ GeV, 
the code is well-equipped for the latest data analyses at present 
accelerators and to become an important tool 
for data-fitting in fermion pair production at a future LC.
   
In the massive case, the code has to be upgraded for the still missing 
branch of continuum top quark pair production at a LC. 
First attempts in this direction have been presented here.
With the upgraded code, or a condensed version
{\tt topfit} for continuum $t\bar{t}$ physics, 
a determination of the weak couplings of the top quark 
may be studied. {\tt ZFITTER}, 
or {\tt topfit} respectively, would then describe all fermion 
pair final-states including radiative corrections and 
different realistic cuts for the 
complete range of energies e.g.~at DAPHNE, LEP/SLC, and at a LC, 
i.e.~for $\sqrt{s}\approx O(1\,\mbox{GeV})$ up to $O(1\,\mbox{TeV})$.
However, at DAPHNE energies a specialized code might be better suited in 
view of the different hadronic final states there which are observed besides 
fermion pairs \cite{Binner:1999bt,Hoefer:2000}.

%
%

\providecommand{\href}[2]{#2}

\end{document}